\documentclass[reprint,twocolumn,superscriptaddress,prl,showpacs,amsmath,amssymb,aps]{revtex4-1}

\usepackage{natbib}	
\usepackage{graphicx,color,rotating}
\usepackage[latin1]{inputenc}
\usepackage{textcomp}
\usepackage{dcolumn}
\usepackage{bm}     
\usepackage{upgreek}
\usepackage{subdepth}  			
\usepackage[latin1]{inputenc}

\usepackage{array}
\newcolumntype{L}[1]{>{\raggedright\let\newline\\\arraybackslash\hspace{0pt}}m{#1}}
\newcolumntype{C}[1]{>{\centering\let\newline\\\arraybackslash\hspace{0pt}}m{#1}}
\newcolumntype{R}[1]{>{\raggedleft\let\newline\\\arraybackslash\hspace{0pt}}m{#1}}

\newcommand{\mr}[1]{\ensuremath{\mathrm{#1}}}
\renewcommand{\vec}[1]{\bm{#1}}
\newcommand{\ee}{\mathrm{e}}
\newcommand{\ii}{\mathrm{i}}
\newcommand{\dm}{\mathrm{d}}

\newcommand{\avr}[1]{\big\langle #1 \big\rangle}

\newcommand{\iot}{{\ii\omega t}}

\newcommand{\pp}{\partial}

\newcommand{\nablabf}{\boldsymbol{\nabla}}

\newcommand{\rot}{\nablabf\times}
\newcommand{\divop}{\nablabf\cdot}
\newcommand{\ppt}{\partial_t}

\newcommand{\een}{\vec{e}}

\newcommand{\fffac}{\vec{f}_\mathrm{ac}}
\newcommand{\fffacI}{\vec{f}_\mathrm{ac}^{(1)}}

\newcommand{\gvec}{\vec{g}}

\newcommand{\rrr}{\vec{r}}

\newcommand{\vvv}{\vec{v}}

\newcommand{\Eac}{E_\mathrm{ac}}

\newcommand{\etaO}{\eta_0}
\newcommand{\etaOO}{\eta_0^{(0)}}

\newcommand{\nuO}{\nu_0}

\newcommand{\phiR}{\phi_\rho}

\newcommand{\kapO}{\kappa_0}
\newcommand{\kapOO}{\kappa_0^{(0)}}

\newcommand{\pO}{p_0}
\newcommand{\pa}{p_\mathrm{a}}

\newcommand{\pI}{p_1}
\newcommand{\pIO}{p_1^{(0)}}

\newcommand{\pII}{p_2}

\newcommand{\vvvI}{\vvv_1}
\newcommand{\vvvIO}{\vvv_1^{(0)}}

\newcommand{\rhoO}{\rho_0}
\newcommand{\rhoOO}{\rho_0^{(0)}}

\newcommand{\rhoI}{\rho_1}
\newcommand{\rhoII}{\rho_2}

\newcommand{\SImum}{\textrm{\textmu{}m}}

\newcommand{\SIms}{\textrm{ms}}
\newcommand{\SImus}{\textrm{\textmu{}s}}

\newcommand{\beq}[1]{\begin{equation} \eqlab{#1}}
\newcommand{\eeq}{\end{equation}}
\newcommand{\bsub}{\begin{subequations}}
\newcommand{\esub}{\end{subequations}}
\def\bal#1\eal{\begin{align}#1\end{align}}
\def\bsubal#1\esubal{\bsub \begin{align}#1\end{align} \esub}
\newcommand{\nn}{\nonumber}
\newcommand{\eqlab}[1]{\label{eq:#1}}
\renewcommand{\eqref}[1]{Eq.~(\ref{eq:#1})}

\newcommand{\eqrefnoEq}[1]{(\ref{eq:#1})}
\newcommand{\eqsref}[2]{Eqs.~(\ref{eq:#1}) and~(\ref{eq:#2})}

\newcommand{\figref}[1]{Fig.~\ref{fig:#1}}
\newcommand{\figsref}[2]{Figs.~\ref{fig:#1} and~\ref{fig:#2}}
\newcommand{\figlab}[1]{\label{fig:#1}}

\newcommand{\lan}{\left\langle}
\newcommand{\ran}{\right\rangle}

\newcommand{\Pibf}{\bm{\Pi}}
\newcommand{\sigmabf}{\bm{\sigma}}

\begin{document}

\title{The Acoustic Force Density Acting on Inhomogeneous Fluids in Acoustic Fields}

\author{Jonas T. Karlsen}
\email{jonkar@fysik.dtu.dk}
\affiliation{Department of Physics, Technical University of Denmark, DTU Physics Building 309, DK-2800 Kongens Lyngby, Denmark}

\author{Per Augustsson}
\affiliation{Department of Biomedical Engineering, Lund University, Ole R\"{o}mers v\"{a}g 3, 22363, Lund, Sweden}

\author{Henrik Bruus}
\email{bruus@fysik.dtu.dk}
\affiliation{Department of Physics, Technical University of Denmark, DTU Physics Building 309, DK-2800 Kongens Lyngby, Denmark}

\date{22 April 2016}

\begin{abstract}
We present a theory for the acoustic force density acting on inhomogeneous fluids in acoustic fields on time scales that are slow compared to the acoustic oscillation period. The acoustic force density depends on gradients in the density and compressibility of the fluid. For microfluidic systems, the theory predicts a relocation of the inhomogeneities into stable field-dependent configurations, which are qualitatively different from the horizontally layered configurations due to gravity. Experimental validation is obtained by confocal imaging of aqueous solutions in a glass-silicon microchip.
\end{abstract}




\maketitle


The physics of acoustic forces on fluids and suspensions has a long and rich history including early work on fundamental phenomena such as acoustic streaming~\cite{LordRayleigh1884, Schlichting1932, Eckart1948, Nyborg1958}, the acoustic radiation force acting on a particle~\cite{King1934, Yosioka1955} or an interface of two immiscible fluids~\cite{Hertz1939}, and acoustic levitation~\cite{Bucks1933, Hanson1964}. Driven by applications related to particle and droplet handling, the field continues to be active with recent advanced studies of acoustic levitators~\cite{Foresti2013, Foresti2014, Marzo2015}, acoustic tweezers and tractor beams~\cite{Courtney2014, Baresch2016, Demore2014}, thermoviscous effects~\cite{Rednikov2011, Muller2014, Karlsen2015}, and in general rapid advances within the field of microscale acoustofluidics~\cite{Bruus2011c}. In the latter, acoustic radiation forces are used to confine, separate, sort or probe particles such as microvesicles~\cite{Evander2015, Lee2015}, cells~\cite{Petersson2007, Wiklund2012b, Collins2015, Ahmed2016, Guo2016}, bacteria~\cite{Hammarstrom2012, Carugo2014}, and biomolecules \cite{Sitters2015}. Biomedical applications include early detection of circulating tumor cells in blood~\cite{Augustsson2012, Li2015} and diagnosis of bloodstream infections~\cite{Hammarstrom2014a}.

The theoretical treatment of acoustic forces involves nonlinear models including multiple length and time scales~\cite{Hamilton2008}. Steady acoustic streaming~\cite{Riley2001} describes a steady fluid motion, spawned by fast-time-scale acoustic dissipation in either boundary layers~\cite{Schlichting1932} or in the bulk~\cite{Eckart1948}. Similarly, the acoustic radiation force acting on a particle~\cite{Karlsen2015} or an interface of two immiscible fluids~\cite{Marr-Lyon2001, Bertin2012} is due to interactions between the incident and the scattered acoustic waves. This force derives from a divergence in the time-averaged momentum-flux-density tensor, which is non-zero only at the position of the particle or the interface.

Recently, in microchannel acoustofluidics experiments, it was discovered that acoustic forces can relocate inhomogeneous aqueous salt solutions and stabilize the resulting density profiles against hydrostatic pressure gradients~\cite{Deshmukh2014}. Building on this discovery, iso-acoustic focusing was subsequently introduced as an equilibrium cell-handling method that overcomes the central issue of cell-size dependency in acoustophoresis~\cite{Augustsson2016}. The method can be considered a microfluidic analog to density gradient centrifugation, achieving spatial separation of different cell-types based on differences in their acousto-mechanical properties. Not surprisingly, the subtle nonlinear acoustic phenomenon of relocation and stabilization of inhomogeneous fluids was discovered in the realm of microfluidics, where typical hydrostatic pressure differences ($\sim1$~Pa) are comparable to, or less than, the acoustic energy densities (1-100~Pa) obtained in typical microchannel resonators~\cite{Barnkob2010,Augustsson2011,Augustsson2016}.

The main goal of this Letter is to provide a theoretical explanation of this phenomenon. To this end, we extend acoustic radiation force theory beyond the requirement of immiscible phases, and present a general theory for the time-averaged acoustic force density acting on a fluid with a continuous spatial variation in density and compressibility.  The starting point of our treatment is to identify and exploit the separation in time scales between the fast time scale of acoustic oscillations and the slow time scale of the oscillation-time-averaged fluid motion. We show that gradients in density and compressibility result in a divergence in the time-averaged momentum-flux-density tensor, which, in contrast to the case of immiscible phases, is generally non-zero everywhere in space. Our theory explains the observed relocation and stabilization of inhomogeneous fluids. Further, we present experimental validation of our theory obtained by confocal imaging in an acoustofluidic glass-silicon microchip.

\textit{Characteristic time scales.---} Consider the sketch in \figref{fig_01} of a long, straight microchannel of cross-sectional width $W = 375~\SImum$ and height $H = 150~\SImum$ filled with a fluid of inhomogeneous density $\rhoO(\rrr)=[1+\delta(\rrr)]\rhoOO$, compressibility $\kapO(\rrr)$, and dynamic viscosity $\etaO(\rrr)$. Here, $\delta(\rrr)$ is the relative deviation away from the reference density $\rhoOO$. Assuming an acoustic standing half-wave resonance at angular frequency $\omega$, the wave number is $k = \omega/c = \pi / W $, where $c=1/\sqrt{\rhoO \kapO}$ is the speed of sound. In terms of the parameters of the microchannel and of water at ambient conditions, the fast acoustic oscillation time scale $t$ is
 \bal
 t \sim \frac{1}{\omega} = \frac{1}{k c} = \frac{1}{\pi} W \sqrt{\rhoO \kapO} \sim 0.1~\SImus .
 \eal
In contrast, the time scales associated with flows driven by hydrostatic pressure gradients are much slower. Given the length scale $H$, the gravitational acceleration $g$, and the kinematic viscosity  $\nuO=\etaO/\rhoO$, we estimate the time scale of inertia $t_\mr{inertia} \sim \sqrt{H/(g\delta)}$, of viscous relaxation $t_\mr{relax} \sim H^2/\nuO$, and of steady shear motion $t_\mr{shear} = \nuO/(Hg\delta)$, the latter obtained by balancing the shear stress $\etaO/t_\mr{shear}$ with the hydrostatic pressure difference $H \rhoO g \delta$. Remarkably, in our system with $\delta \approx 0.1$, all time scales are of order 10~ms, henceforth denoted the slow time scale $\tau$,
 \bal
 \tau \sim  t_\mr{inertia} \sim t_\mr{relax} \sim t_\mr{shear} \sim 10~\SIms.
 \eal
Furthermore, for $\Eac \sim \rhoO g H$ the time scale of flows driven by time-averaged acoustic forces is also $\tau$. Hence, we have identified a separation of time scales into a fast acoustic time scale $t$ and a slow time scale $\tau \sim 10^5 t$. This separation is sufficient to ensure $\tau \gg t$ in general, even for large variations in parameter values.

\begin{figure}[!t]
\centering
\includegraphics[width=1.0\columnwidth]{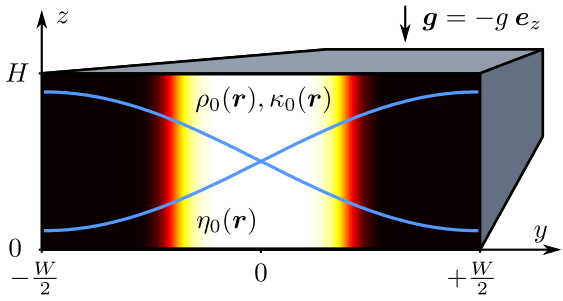}
\caption[]{\figlab{fig_01} (color online)
Sketch of a long, straight acoustofluidic microchannel of length $L = 40$~mm along $x$, width $W=375~\SImum$ and height $H=150~\SImum$ with an imposed half-wave acoustic pressure resonance (sinusoidal curves) inside a glass-silicon chip. A salt concentration (color scale: black low, white high) leads to an inhomogeneous fluid density $\rhoO(\rrr)$, compressibility $\kapO(\rrr)$, and dynamic viscosity $\etaO(\rrr)$. The gravitational acceleration is $\gvec=-g \een_z$.}

\end{figure}

\textit{Fast time-scale dynamics.---} The dynamics at the fast time scale $t$ describes acoustics for which viscosity may be neglected~\cite{Landau1993, Morse1986, Bruus2012}. On this time scale $\rhoO$, $\kapO$, and $\etaO$ can be assumed stationary, and the acoustic fields are treated as time-harmonic perturbations at the angular frequency $\omega$~\cite{Bruus2012}. The perturbation expansion for the density $\rho$ thus takes the form
 \beq{PertExpansion}
 \rho = \rhoO(\rrr,\tau) + \rhoI(\rrr,\tau)\:\ee^{-\iot} + \rhoII(\rrr,t,\tau) ,
 \eeq
and likewise for the pressure $p$ and the velocity $\vvv$. In terms of the material derivative $\frac{\dm}{\dm t} = \ppt +  (\vvv\cdot\nablabf)$, the density-pressure relation for a fluid particle is
 \beq{DensityPressureRelation}
 \frac{\dm \rho}{\dm t} =
 \rhoO \kapO \frac{\dm p}{\dm t},\;
 \text{ where by definition } \;
 \frac{1}{c^2} = \rhoO\kapO.
 \eeq
Here, $c$ is the local speed of sound, which depends on position through the inhomogeneity in $\kapO$ and $\rhoO$. Combining \eqsref{PertExpansion}{DensityPressureRelation} leads to the first-order relation
 \beq{DensityPressureFirstOrder}
 \ppt \rhoI + (\vvvI\cdot\nablabf) \rhoO = \rhoO \kapO \Big[ \ppt \pI + (\vvvI\cdot\nablabf) \pO  \Big].
 \eeq
From the governing equations for mass and momentum~\cite{Landau1993, Morse1986, Bruus2012} it follows that $\nablabf\pO \ll c^2\nablabf \rhoO$, and the term involving $\nablabf\pO$ in \eqref{DensityPressureFirstOrder} is negligible. This results in the first-order equations,
 \bsub
 \eqlab{FirstOrderEqs}
 \bal
 \eqlab{FirstOrderMass}
 \kapO \ppt \pI &= -\nablabf\cdot\vvvI, \\
 \eqlab{FirstOrderMomentum}
 \rhoO \ppt \vvvI &= -\nablabf \pI,
 \eal
 \esub
and the wave equation for the acoustic pressure $\pI$ in an inhomogeneous fluid~\cite{Morse1986, Pierce1990},
 \beq{inhomogWaveEq}
 \frac{1}{c^2} \pp^2_{t} \pI = \rhoO \nablabf\cdot\bigg[\frac{1}{\rhoO}\nablabf\pI\bigg].
 \eeq
As will be used later, the rotation of \eqref{FirstOrderMomentum} leads to $\rot(\rhoO\vvvI)=\vec{0}$, which implies that acoustics in inhomogeneous fluids should be formulated in terms of the mass current potential $\phiR$ instead of the usual velocity potential,
 \beq{phiR}
 \rhoO\vvvI = \nablabf\phiR, \quad \text{and} \quad
 \pI = -\ppt \phiR.
 \eeq
Combining \eqsref{FirstOrderMass}{phiR} reveals that the mass current potential $\phiR$ fulfills the same wave equation as $\pI$.

\textit{The acoustic force density.---} The first-order acoustic fields lead to no net fluid displacement, since the time-average $\avr{g_1} = \frac{1}{T}\int_0^T g_1\:\dm t$ over one oscillation period $T$ of any time-harmonic first-order field $g_1$ is zero. The description of time-averaged effects thus requires the solution of the time-averaged second-order equations, and the introduction of the time-averaged momentum-flux-density tensor $\avr{\Pibf}$~\cite{Landau1993},
 \beq{Pibf}
 \avr{\Pibf} = \avr{\pII} \mathbf{I} +  \avr{ \rhoO \vvvI \vvvI }.
 \eeq
Here, $\mathbf{I}$ is the unit tensor, and the second-order mean Eulerian excess pressure $\lan \pII \ran$ is given by the difference between the time-averaged acoustic potential and kinetic energy densities~\cite{Lee1993, Smith2001, Zhang2011},
 \beq{p2}
 \avr{\pII}
 = \frac12 \kapO \avr{|\pI|^2} - \frac12 \rhoO \avr{ |\vvvI|^2 }.
 \eeq
In the well-known case of a particle suspended in a homogeneous fluid in an acoustic field, the deviation in density and compressibility introduced by the particle leads to a scattered acoustic wave, which induces a divergence $\divop\avr{\Pibf}$ in $\avr{\Pibf}$. The radiation force exerted on the particle may then be obtained by integrating the force density $-\divop\avr{\Pibf}$ over a volume enclosing the particle, thereby picking out the divergence at the particle position~\cite{Karlsen2015, Fan2008}.

In the case of an inhomogeneous fluid, the gradient in the continuous material parameters $\rhoO(\rrr)$ and $\kapO(\rrr)$ will likewise lead to a non-zero divergence in $\avr{\Pibf}$. This is the origin of the acoustic force density $\fffac$ acting on the inhomogeneous fluid at the slow time scale. Consequently, we introduce $\fffac$ as
 \beq{facDef}
 \fffac = -\divop\avr{\Pibf} = - \nablabf\avr{\pII} - \divop\avr{\rhoO\vvvI\vvvI} .
 \eeq
Here, $\avr{\pII}$ is given by the local expression~\eqrefnoEq{p2}, which remains true in an inhomogeneous fluid, while the divergence term is rewritten using~\eqref{FirstOrderMass} for $\divop\vvvI$ and \eqref{phiR} defining the mass current potential $\phiR$,
 \bsub
 \eqlab{divMom}
 \bal
 & \divop\avr{\rhoO\vvvI\vvvI} \nn \\
 &= \avr{\vvvI\cdot\nablabf(\rhoO\vvvI)} + \avr{\rhoO\vvvI(\divop\vvvI)} , \\
 &= \Big\langle \Big(\frac{1}{\rhoO}\nablabf\phiR\Big)\cdot\nablabf(\nablabf\phiR) \Big\rangle
 + \avr{(\nablabf\phiR)(\kapO\pp^2_{t}\phiR)} , \\
 \eqlab{divMomEqC}
 &= \frac{1}{2\rhoO}\nablabf\avr{|\nablabf\phiR|^2} - \kapO\avr{(\nablabf\ppt\phiR)(\ppt\phiR)} , \\
 &= \frac{1}{2\rhoO}\nablabf\avr{|\nablabf\phiR|^2} - \frac{1}{2}\kapO\nablabf\avr{|\ppt\phiR|^2} , \\
 \eqlab{divMomEqE}
 &= \frac{1}{2\rhoO}\nablabf \avr{|\rhoO\vvvI|^2} - \frac12 \kapO\nablabf\avr{|\pI|^2} .
 \eal
 \esub
In \eqref{divMomEqC} we have used $\avr{f_1 (\pp_t g_1)} = -\avr{(\pp_t f_1) g_1}$, valid for time-harmonic fields $f_1$ and $g_1$.

Combining Eqs.~\eqrefnoEq{Pibf}-\eqrefnoEq{divMom} and evaluating the time averages~\footnote{{The time average of the product of two time-harmonic complex-valued fields $f_1$ and $g_1$ is $\langle f_1\:g_1 \rangle = \frac12 \mathrm{Re}[f_1^* g_1]$, where the asterisk denotes complex conjugation.}}, we arrive at our final expression for the acoustic force density $\fffac$ acting on an inhomogeneous fluid,
 \beq{facFinal}
 \fffac = - \frac14 |\pI|^2 \nablabf\kapO - \frac14 |\vvvI|^2 \nablabf\rhoO .
 \eeq
This main result, obtained in part by using the mass current potential $\phiR$, demonstrates that gradients in compressibility and density lead to a time-averaged acoustic force density acting on an inhomogeneous fluid. This force density is generally non-zero in every point in space, in contrast to the cases of a localized interface or a particle.

\textit{Analytical approximation for $|\delta| \ll 1$.---}
We can obtain analytical results that provide physical insight in the experimentally relevant limit of fluids with a constant speed of sound $c$ and a weakly varying density~\cite{Deshmukh2014, Augustsson2016}. Writing the latter as $\rhoO(\rrr,\tau) = \rhoOO[1 + \delta(\rrr,\tau)]$, where $|\delta(\rrr,\tau)| \ll 1$ and the superscript $(0)$ indicates zeroth-order in $\delta$,
we obtain $\nablabf\kapO=\frac{1}{c^2} \nablabf(\frac{1}{\rhoO})=-\frac{\kapO}{\rhoO} \nablabf \rhoO$. To first order in $\delta$, $\fffac$ in \eqref{facFinal} thus becomes,
 \bal
 \eqlab{facWeakInhom}
 \fffacI = \Big[ \frac14 \kapOO |\pIO|^2 - \frac14 \rhoOO |\vvvIO|^2 \Big] \nablabf\delta.
 \eal
Compared to \eqref{facFinal}, this expression constitutes  a major simplification, since it is linear in $\nablabf\delta$ and it employs the $\delta$-independent homogeneous-fluid fields $\pIO$ and $\vvvIO$.

Based on \eqref{facWeakInhom}, we demonstrate analytically that our theory is capable of explaining recent experimental results~\cite{Deshmukh2014, Augustsson2016}. For the system in~\figref{fig_01}, with a horizontal acoustic half-wave pressure resonance of amplitude $\pa$, the homogeneous-fluid field solution takes the form,
 \bsub
 \eqlab{fieldsHalfWave}
 \bal
 \pIO &= \pa\;\sin(k y), \quad \text{with} \quad k = \frac{\pi}{W},\\
 \vvvIO &= \frac{\pa}{\ii\rhoOO c}\:\cos(ky)\:\een_y .
 \eal
 \esub
In this case \eqref{facWeakInhom} reduces to
 \beq{facHalfWave}
 \fffacI = -\cos(2 k y)\:E^{(0)}_\mathrm{ac}\:\nablabf \delta ,
 \eeq
where $E^{(0)}_\mathrm{ac}=\frac{1}{4}\kapOO\pa^2$ is the homogeneous-fluid time-averaged acoustic energy density. Consider a fluid that is initially stratified in horizontal density layers $\delta(\rrr,0)=\delta(z)$ (not vertical layers as in \figref{fig_01}), with the dense fluid occupying the floor of the channel ($\pp_z \delta < 0$). Equation~\eqrefnoEq{facHalfWave} then predicts that the fluid layers will be pushed downwards near the channel sides, but upwards in the center. This explains the initial phase in the slow-time-scale relocation of the denser fluid to the center of the channel observed experimentally~\cite{Deshmukh2014}.

\textit{Slow time-scale dynamics.---} Our experiments confirm the observation~\cite{Augustsson2016} that acoustic streaming is suppressed in the bulk of an inhomogeneous fluid. On the slow time scale $\tau$, the dynamics is therefore governed by the acoustic force density $\fffac$, the gravitational force density $\rhoO\gvec$, and the induced viscous stress, such that the Navier--Stokes equation and the continuity equation take the form
 \bsub
 \eqlab{DynamicsSlow}
 \bal
 \eqlab{NSSlow}
 \pp_\tau (\rhoO \vvv) &= \divop \big[ \sigmabf - \rhoO\vvv\vvv \big] + \fffac + \rhoO \gvec , \\
 \eqlab{ContSlow}
 \pp_\tau \rhoO &= - \divop \big( \rhoO \vvv \big) ,
 \eal
where $\sigmabf$ is the stress tensor, given by
 \bal
 \sigmabf = - p \: \mathbf{I} + \etaO \Big[ \nablabf \vvv + (\nablabf \vvv)^\mathrm{T} \Big] + \Big(\etaO^\mathrm{b} - \frac{2}{3}\etaO \Big) (\divop\vvv) \: \mathbf{I} . \nn
 \eal
Here, the superscript "T" indicates tensor transposition and $\etaO^\mathrm{b}$ is the bulk viscosity, for which we use the value of water~\cite{Muller2014}. The inhomogeneity in the fluid parameters is assumed to be caused by a spatially varying concentration field $s(\rrr,\tau)$ of a solute molecule with diffusivity $D$, satisfying the advection-diffusion equation,
 \bal
 \eqlab{DiffusionSlow}
 \pp_\tau s = - \divop \big[ - D \nablabf s + \vvv s \big] .
 \eal
 \esub

In our experimental setup, aqueous solutions of iodixanol are used to create inhomogeneities in density, while maintaining an approximately constant speed of sound. The relevant solution properties have been measured as functions of the iodixanol volume-fraction concentration $s$ in our previous work~\cite{Augustsson2016}. For the density $\rhoO$ and viscosity $\etaO$, the resulting fits are $\rhoO=\rhoOO[1+ a_1 s]$ and $\etaO=\etaOO[1+b_1 s + b_2 s^2 + b_3 s^3]$, with $\rhoOO=1005$~kg/m$^3$, $\etaOO=0.954$~mPa$\,$s, and $a_1=0.522$, $b_1=2.05$, $b_2=2.54$, $b_3=22.8$. The diffusivity was measured \emph{in situ} to be $D=0.9\times10^{-10}$~$\mathrm{m^2/s}$.

\begin{figure}[!t]
\centering
\includegraphics[width=1.0\columnwidth]{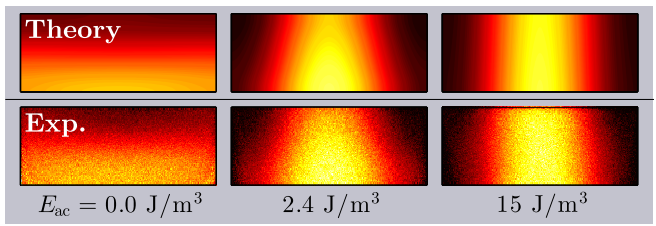}
\caption[]{\figlab{fig_02} (color online)
Simulation of the theory (top) and experimental confocal image (bottom) of the cross-sectional concentration of iodixanol after 17~s retention time for three acoustic energy densities $\Eac$. Initially, the denser fluorescently marked fluid (36\% iodixanol, white) is in the center and the less dense fluid (10\% iodixanol, black) is at the sides, see \figref{fig_01}. The stable configurations confirm the observation in Ref.~\cite{Augustsson2016} that acoustic streaming is suppressed in inhomogeneous fluids. There are no free fitting parameters.}
\end{figure}

\textit{Comparison to experiments.---} Our experimental setup is described in detail in Ref.~\cite{Augustsson2016}. The microchannel in the glass-silicon microchip has the dimensions given in \figref{fig_01}. The horizontal half-wave resonance is excited by driving an attached piezo transducer with an ac voltage $U$ swept repeatedly in frequency from 1.9~MHz to 2.1~MHz in cycles of 1~ms to ensure stable operation. The resulting average acoustic energy density is measured by observing the acoustic focusing of $5~\SImum$ beads~\footnote{{The average acoustic energy density $\Eac$ is estimated as a function of the piezo-transducer voltage $U$ by observing the focusing of 5~$\upmu$m polystyrene particles in a homogeneous 10\% iodixanol solution and comparing to theoretical models, see Refs.~\cite{Barnkob2010, Augustsson2011}. In our system, this yields $\Eac=k U^2$ with $k=1.2$~$\mathrm{J\,m^{-3}\,V^{-2}}$~\cite{Augustsson2016}.}}. The channel inlet conditions are illustrated in \figref{fig_01}: a fluorescently marked 36\% iodixanol solution (white) is laminated by 10\% iodixanol solutions on either side (black)~\footnote{{We use a fluorescent dextran tracer with the molecular weight 3000~Da and a diffusivity close to that of iodixanol, which allows indirect visualization of the iodixanol concentration profile, see Ref.~\cite{Augustsson2016}.}}. The corresponding density variation is 13\% with the maximum at the channel center. At the outlet, after a retention time of $\tau_\mathrm{ret}=17$~s, the fluorescence profile is imaged using confocal microscopy in the channel cross section. The characteristic time for diffusion across one third of the channel width is $\tau_\mathrm{diff}=\frac{1}{2D}\big(\frac{W}{3}\big)^2=87$~s, so diffusion is important but not dominant in the experiment.

We simulate numerically the time evolution in the system using the finite-element solver COMSOL Multiphysics~\cite{COMSOL52}, by implementing \eqsref{facHalfWave}{DynamicsSlow} with the measured dependencies of density $\rhoO(s)$ and viscosity $\etaO(s)$ on concentration $s$. The initial concentration field $s(\rrr,0)$ is set to the inlet conditions allowing the concentration field $s(\rrr,\tau_\mathrm{ret})$ to be compared to the experimental images. The acoustic energy density $\Eac$ entering the model is set to the measured experimental value, which leaves no free parameters. Concerning the validity of the numerical solutions, several convergence tests were performed \cite{Muller2014}, and the integral of $s$ over the domain was conserved in time with a relative error of order $10^{-3}$.

In \figref{fig_02} we compare the numerically simulated and experimentally measured concentration fields $s(\rrr,\tau)$ at time $\tau_\mathrm{ret}=17$~s for three acoustic energy densities $\Eac$. For $\Eac=0~\mathrm{J/m^3}$ the initially vertical center fluid column of high density (\figref{fig_01}, white) has collapsed and relocated to the channel bottom due to gravity. For $\Eac = 15~\mathrm{J/m^3}$, the acoustic force density stabilizes the denser vertical fluid column against gravity, such that it broadens only by diffusion. For the intermediate value $\Eac = 2.4~\mathrm{J/m^3}$, where the gravitational and acoustic forces are comparable, the stable configuration has a triangular shape. Note that the good agreement between the simulated and measured concentration profiles has been obtained without free fitting parameters.

In~\figref{fig_03} we show time-resolved simulations obtained with $\Eac=10$~$\mathrm{J/m^3}$ for (a) the stable initial configuration with the denser fluid at the center, (b) the unstable initial configuration with the denser fluid at the bottom, and (c) the unstable initial configuration with the denser fluid at the sides. While the stable initial configuration~(a) evolves only by diffusion, the unstable initial configurations~(b) and~(c) evolve by complex advection patterns into essentially the same stable configuration with the denser fluid at the center. This fluid relocation is in full qualitative agreement with experiments~\cite{Deshmukh2014}.

\begin{figure}[!t]
\centering
\includegraphics[width=1.0\columnwidth]{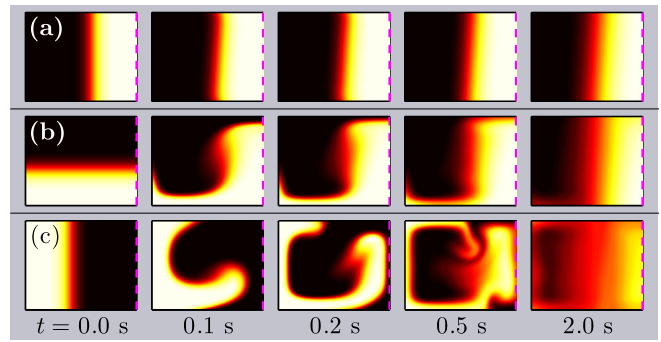}
\caption[]{\figlab{fig_03} (color online)
Simulation for $\Eac=10$~$\mathrm{J/m^3}$ of the time-evolution of the iodixanol concentration profile in the vertical $yz$ plane symmetric around $y=0$ (dashed line), with only the left half $-\frac12 W \leq y\leq 0$ shown. Three different initial configurations of the dense (36\% iodixanol, white) and less dense (10\% iodixanol, black) solution give rise to different time evolutions. (a) A vertical slab of the dense fluid in the center. (b) A horizontal slab of the dense fluid at the bottom. (c) Two vertical slabs of the dense fluid at the sides. All configurations develop towards a stable configuration with the dense fluid located as a nearly vertical slab in the center.}
\end{figure}

\textit{Discussion.---} Our theory for the acoustic force density acting on an inhomogeneous fluid explains recent experimental observations~\cite{Deshmukh2014, Augustsson2016}, and agrees with our experimental validation without free parameters. The additional observation that bulk streaming is absent in inhomogeneous fluids~\cite{Augustsson2016} has not been treated in this Letter. However, \figsref{fig_02}{fig_03} demonstrate that the acoustic force density stabilizes a particular inhomogeneous configuration, and this is likely to also explain the suppression of streaming. By adding acoustic boundary layers to our model, we are currently investigating this hypothesis. The extension of acoustic radiation force theory to include inhomogeneous fluids through the introduction of the acoustic force density~\eqrefnoEq{facFinal} represents an increased understanding of acoustofludics in general, and further has the potential to open up new ways for microscale handling of fluids and particles using acoustic fields.

We thank Mads Givskov Senstius, Technical University of Denmark, for assistance with experiments. P.A. had financial support from the Swedish Research Council (grant no.\ 2012-6708), the Royal Physiographic Society, and the Birgit and Hellmuth Hertz' Foundation.

%
%


%

\end{document}